\documentclass[%
reprint,
superscriptaddress,
 amsmath,amssymb,
 prl,
]{revtex4-1}

\usepackage{graphicx}
\usepackage{dcolumn}
\usepackage{bm}
\usepackage{float}

\begin{document}


\title{Quantum multifractality in thermal conduction across random interfaces}

\author{Taishan Zhu}
\affiliation{%
Department of Materials Science and Engineering,
Massachusetts Institute of Technology,
Cambridge, MA 02139, USA
}
\affiliation{%
Research Laboratory of Electronics,
Massachusetts Institute of Technology,
Cambridge, MA 02139, USA
}%


\author{Giuseppe Romano}
\affiliation{
Institute for Soldiers Nanotechnologies,
Massachusetts Institute of Technology,
Cambridge, MA 02139, USA
}%
\author{Lina Yang}
\affiliation{%
School of Aerospace Engineereing,Beijing Institute of Technology, Beijing, 100081, China
}%



\author{Martin Ostoja-Starzewski}
\affiliation{
Department of Mechanical Science and Engineering}
\affiliation{
Beckman Institute and Institute for Condensed
Matter Theory, University of Illinois at Urbana-Champaign, Urbana, IL 61801, USA
}

\author{Jeffrey C. Grossman}
\email{Corresponding author: jcg@mit.edu}
\affiliation{Department of Materials Science and Engineering,
Massachusetts Institute of Technology,
Cambridge, MA 02139, USA
}


\begin{abstract}
Self-affine morphology of random interfaces governs their functionalities across tribological, geological, (opto-)electrical and biological applications. 
However, the knowledge of how energy carriers or generally classical/quantum waves interact with structural irregularity
is still incomplete.
In this work, 
we study vibrational energy transport through random interfaces exhibiting different correlation functions on the two-dimensional hexagonal lattice.
We show that random interfaces at the atomic scale are Cantor composites populated on geometrical fractals, thus multifractals, and calculate their quantized conductance using atomistic approaches.
We obtain a universal scaling law, which contains self-similarity for mass perturbation, and exponential scaling of structural irregularity quantified by fractal dimension.
The multifractal nature and Cantor-composite picture may also be extendable to charge and photon transport across random interfaces.
\end{abstract}

\maketitle


\section{Introduction}

Interfaces enable sharp potential change in a confined space and plays key roles in various physical, biological, and geological phenomena. \cite{ogilvy1987wave, deak2004vibrational, mandelbrot1983fractal, sahimi1993fractal} 
Depending on the scenario, the potential can be chemical \cite{deak2004vibrational}, electrical \cite{ogilvy1987wave}, and thermal \cite{cahill2003nanoscale,monachon2016thermal,ramiere2016thermal,cui2020emerging,giri2020review}.
For instance, the temperature jump across interfaces, and thus induced thermal resistance, \cite{kapitza1941study} is the driving mechanism for thermoelectrics based on heterostructures and superlattices \cite{sofo1994thermoelectric, broido1995effect,harman2002quantum,vashaee2004improved}. 
In other applications, such as electronics cooling, this interface resistance must be minimized or avoided if possible \cite{pop2006heat}.
Although atomic-scale sharp interfaces can be manufactured \cite{liu2013plane,wen2009formation}, both naturally arising and human-made interfaces are most often random configurationally \cite{cahill2003nanoscale,gotsmann2013quantized}.
The randomness can reside in a mixed layer or can be of multi-asperity type \cite{gotsmann2013quantized, wen2009formation}.
Even for a clearly-cut and perfect-contact interface, its profile can be tortuous,
and it remains a challenge to understand how the structural randomness modifies vibrational energy transport across these interfaces.

\begin{figure*}[htp]
\includegraphics[width=0.70\linewidth]{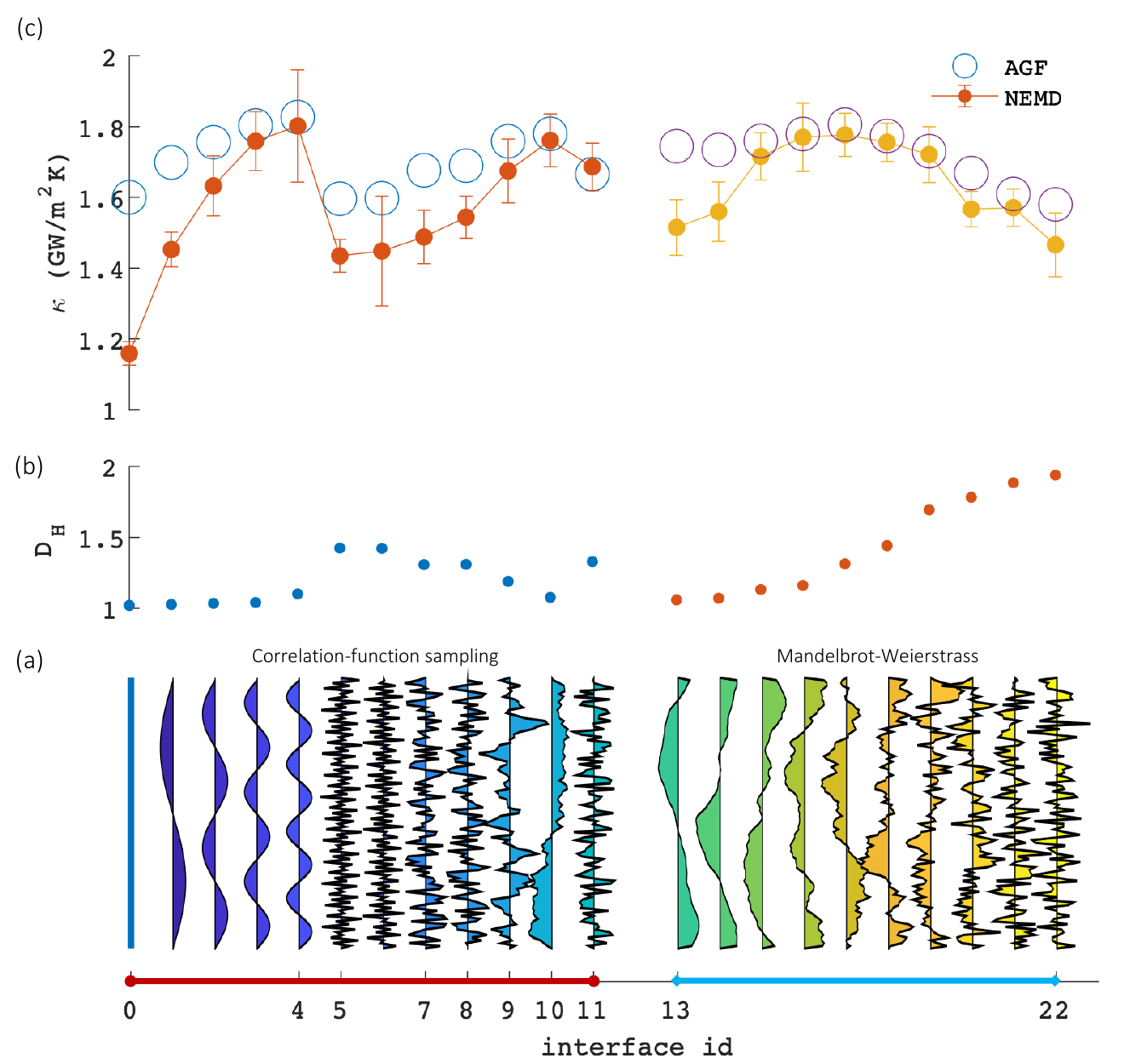}
\caption{\label{figModel} Morphological models generated in real and reciprocal spaces, and their corresponding dimension and conductance. (a) Interfaces No. 0-11 are generated by different correlation functions between two limits: flat (No. 0) and Poissonian (No. 11).  
No. 13-22 are fractals \emph{via} Mandelbrot-Weierstrass approach.
(b) Hausdorf dimension $D_H$ measured by Richardson method for model interfaces in (a). Random interfaces having  $D_H>1$ are fractals. (c) Thermal conductance calculated by atomic Green's function (AGF) and nonequilibrium molecular dynamics (NEMD). 
}
\end{figure*}

Since Kapitza's experiment \cite{kapitza1941study}, two limiting situations have been extensively studied:
i) harmonic flat, and ii) anharmnic Poissonian   (e.g. \cite{swartz1989thermal,ramiere2016thermal,ravichandran2018spectrally}).
For a flat interface, with null randomness, principles of wave optics and acoustics can be applied to derive the energy transmission coefficient, which leads to the acoustic mismatch model (AMM) \cite{little1959transport}.
The transmission is harmonic.
In contrast, for a completely random interface, for which white-noise randomness is assumed, vibrational modes are often hypothesized to be fully thermally equilibriated  adjacent to the interface. 
The latter results in the diffusive mismatch model (DMM) \cite{swartz1989thermal}.
These two models are widely known to formulate the upper and lower limits of interfacial conductance. \cite{monachon2016thermal}
Note that these models are theoretically accurate for gray media, since polarization and its conversion has not been explicitly considered.
Recent works have shown discrepancies between these models and more accurate calculations and experiments (e.g. \cite{zhou2017full, gaskins2018thermal, feng2019unexpected}).


In terms of structure-conductance relation, the challenge of understanding energy transport across random interfaces is two-fold: i) the gap between flat and Poissonian structures is not systematically filled; ii) nonuniqueness. 
For the first aspect, random interfaces have been studied mostly near the two limits in a perturbative manner (e.g., polygons and ``alloy" models \cite{ravichandran2018spectrally, merabia2014thermal, lee2016nanostructures}).
A systematic way to represent a random interface is lacking, which makes it difficult to quantify the difference/similarity between two random interfaces.
Two visually similar interfaces can be rather different in the reciprocal space (see SI).
For the second aspect, multiple morphologically different interfaces can have the same conductance.
However, since the similarity of two random interfaces can not be quantified, it is unpredictable when two random interfaces have similar or different conductance.
Further, universal scaling laws have been found in many random/stochastic phenomena \cite{mandelbrot1983fractal, stanley1987introduction}.
Such scaling laws can help deepen our understanding and shed light on underlying trends in the otherwise apparent randomness.
However, it is unclear if similar universal laws exist in the interfacial conductance of random interfaces.

Another question unanswered by the classic picture is whether the  mesoscale for interfacial conduction exists, and how to define it if it does. \cite{cercignani1975theory}
The classic picture assumes thermal carriers are either completely not or fully thermally equilibriated, 
represented respectively by the AMM and DMM models.
Kinetic-theoretically these correspond to ballistic and diffusive transport regimes,
where the Knudsen number ($Kn$) defined by the ratio of carrier mean free path to the characteristic length of the interface approaches to infinity or zero \cite{chen2005nanoscale, zhu2010theoretical}.
Although interesting physics arises often during the transitional regime (\textit{e.g.}, $0.01<Kn<100$), where the transport of carriers is noncontinuum and nonequilibrium, \cite{zhu2010origin} this regime is not defined yet for interfacial transport.




In this work, we bridge the two limits by i) developing a unified fractal representation generally applicable to model random interfaces, and ii) assessing the universality in their lattice conductance.
We use both deterministic correlation functions and a random Mandelbrot-Weierstrass (MW) approach for geometry construction. 
We deduce a universal scaling law for the conductance of the interfaces generated by different correlations, which we will show is a function of fractal dimension.
Such fractal dimension, as a reverse problem,  can be well controlled by the MW method.
Therefore, we use the interfaces from MW approach to test the universal scaling law proposed for general random interfaces.
Futher, considering the local structures along interfaces, we reveal the multifractal nature of random interfaces at the atomic scale, which defines the mesoscale of lattice thermal conduction.

\section{Rough interfaces are fractals}

We first develop a fractal representation for random interfaces. 
Early experiments demonstrated the morphological complexity of interfaces, 
and culminated in the revelation of structural correlation and self-affinity embedded in randomness. \cite{mandelbrot1983fractal, milanese2019emergence} 
Still, most recent studies on interficial vibrational transport often adopted polygon or ``alloy" models, \cite{merabia2014thermal, lee2016nanostructures, tian2012enhancing, liu2016topological, ravichandran2018spectrally} 
although realistic morphologies could be constructed if structural correlations were considered. \cite{majumdar1991fractal, milanese2019emergence}
More realistic models are the mathematical objects constructed by various iterative 
and/or spectral methods, \cite{mandelbrot1983fractal,falconer2004fractal} but these models have even been less investigated for energy transport.
We will construct random interfaces through two approaches: 
the first is a sampling from given correlation function and the other is a spectral method.

Our first approach samples random interfaces from prescribed correlation functions, which  can be measured by experiments.
Denoting the fluctuations in an interface profile ($\delta(\boldsymbol{x})$), we view a random interface as a random sequence or spatial random walk on the support plane $\boldsymbol{x}$.
In this work, we focus on planar hexagonal lattice, thus $\boldsymbol{x}$ reducing to the set $\{x_i\}$, $i$ being the lattice index.
The structure in $\delta(\boldsymbol{x})$ could be described by the correlation function $\left< \delta(x_i) \delta(x_j) \right>$.
The two limits, flat ($\delta(x_i)=0$) and Poissonian ($\left< \delta(x_i) \delta(x_j) \right>$=0),
are shown in Fig. \ref{figModel} (a) as No. 0 and No. 11.
Here $x$ is the vertical direction.
More correlated functions included are quasi-random uniforms (No. 5-7, respectively Sobol, Halton, and Latin cubic), hyperuniform (No. 8), and exponentially correlated Gaussians (No. 9-10, with correlation length of 5 and 25 lattice constants).
Details of sampling from these correlation functions can be found in, for example, Refs. \cite{torquato2015ensemble, falconer2004fractal}.
To stress the effects of randomness, we also study sinusoidal waves (No. 1-4) with growing wavevector.

Interface profiles No. 13-22 are constructed reversely from given $D_H$, other than a correlation function, by Mandelbrot-Weierstrass (M-W) method. \cite{mandelbrot1983fractal} 
Different from the above real-space sampling, the M-W method sums up component ``subwaves'' in the reciprocal space,
\begin{equation}
    \delta(x)=\sum_{q\in Q} \exp(2\pi \mathrm{i} ~ x \cdot q) \hat{\delta}_q + C.C. \; ,
\end{equation}
where $q$ is the wavevector components drawn from set $Q$, which could be uniformly linear as for Fourier transform $q_{j}=2\pi j/a_0$, $j\in Z$, and $a_0$ a length scale (e.g. lattice constant), or geometrical series $q_{j+1}=c \cdot q_j, c\in [1.1,3]$, or power series $q_{j+1}=q_j^\gamma, \gamma \in [0.7,1.4]$. \cite{mandelbrot1983fractal, falconer2004fractal} 
We choose Fourier series because of its physical significance of wavevectors, 
and the upper/lower limits of $q$ correspond to the lattice constant and interface length of computational samples.
More importantly, it is known that the component power, $\hat{\delta}_q^2 \propto q^{\beta}$ determines the fractal dimension, $D_H=2+\beta/2$. \cite{mandelbrot1983fractal}
With the M-W approach, as shown in Fig. \ref{figModel}(b), we can now cover the regime between white noise ($\beta=0$) and brown noise ($\beta=-2$), the former being equivalent to the above Poissonian limit.

These constructed profiles exhibit varying level of irregularity.
The sinusoidal profiles are smoother than correlated Gaussians and in turn the quasi-random and hyperuniform.
To quantify structural disorder, we calculate their fractal dimensions ($D_H$) through Richardson box method,\cite{mandelbrot1983fractal} as shown in Fig.\ref{figModel}(b).
Except the long-wavelength sinusoidal interfaces (No. 0-3),
all the model profiles 
exhibit fractal dimensions higher than the topological dimension $D_T=1$. 
Therefore, atomic roughness renders all the interfaces other than No. 0-3 to be fractal and self-affine in nature. 
While the correlation-function-based method is a natural option to connect with experimental measurements, the M-W construction is ideal for theoretical studies where $D_H$ can be prescribed {\it a priori}.    
Note that the polygon-shaped interfaces in the literature can also be constructed by both approaches.

\section{Universality \& quantization in $\kappa$}

With the random interfaces generated above, we now turn to their thermal conductance $\kappa$.
It is still unclear how structural disorder along an interface can impact its $\kappa$.
For instance, the AMM/DMM formulation could explain the reduced conductance for structured and random interfaces \cite{merabia2014thermal, lee2016nanostructures}, 
but more recent studies also show the possibility to use structural defects to enhance interfacial conductance \cite{tian2012enhancing, liu2016topological}.
The mechanism for such disorder-mediated enhancement is still under debate.
In order to focus on the structural disorder, we model the interactions between all atoms identically as Tersoff model for graphene, \cite{lindsay2010optimized}
but set the atomic masses to be different at two sides of the interfaces: one side is graphene $m_C=12.0$, and the other side is heavy graphene with atomic masses $m_h=c \, m_C, c=2,4,6,8$, which leads to mass differences
$$
\Delta_m=\frac{m_h-m_C}{m_C} \in \{1,3,5,7 \} \; .
$$
In terms of Anderson theory, disorder can be categorized into diagonal and off-diagonal types. \cite{anderson1958absence}
In our case of lattice transport, mass change corresponds to diagonal disorder, while change in inter-atomic potentials results in both diagonal and off-diagonal disorder.

\begin{figure*}[htp]
\includegraphics[width=0.7 \linewidth]{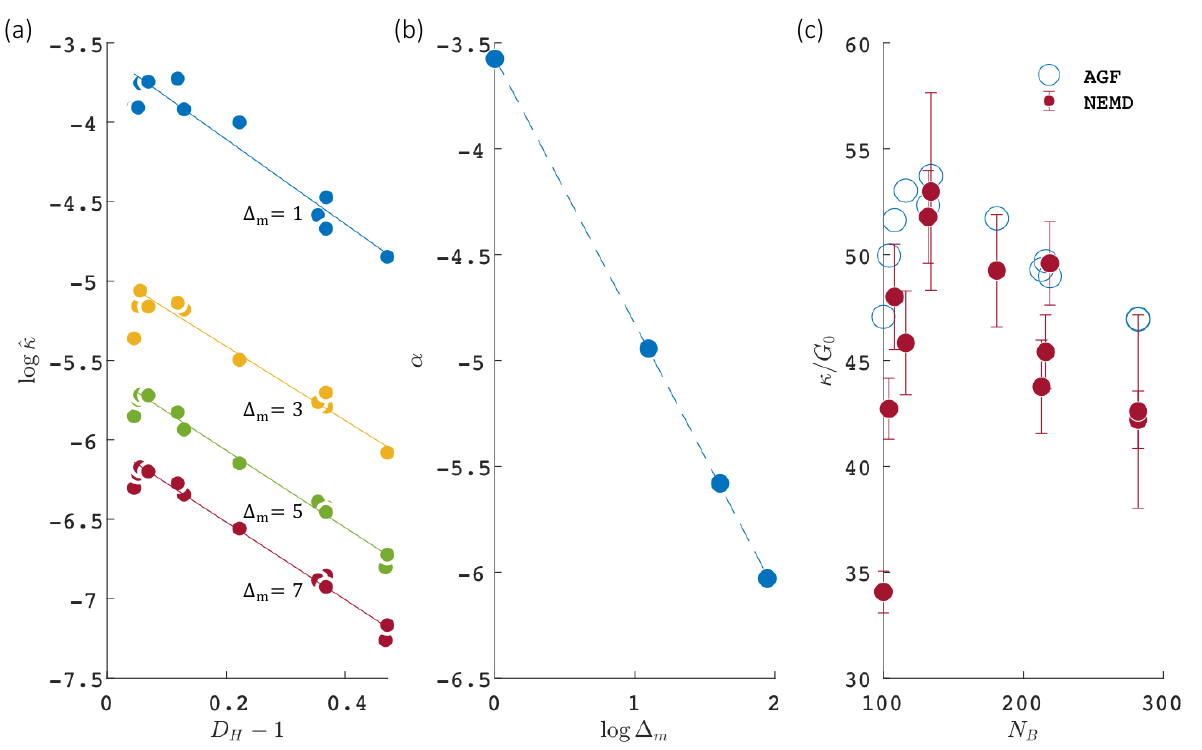}
\caption{\label{fig2} Universality in specific conductance $\hat{\kappa}$. (a) $\hat{\kappa}$ scales linearly with the fractal dimension $D_H$, which suggests an exponential scaling exists. The fitting line is $\log\hat{\kappa}=-(5/2 \pm 0.02) (D_H-1) + \alpha$. (b) the intercept $\alpha$ of the linear scaling in (a) decreases linearly with mass difference $\log\Delta_m$, and the broken fitting line is $\alpha=-(5/4\pm 0.03) \log\Delta_m - (3.5 \pm 0.1)$.
(c) $\hat{\kappa}$ as multiples of universal conductance quantum $G_0$.
}
\end{figure*}

The conductance $\kappa$ of our model interfaces is calculated by atomic Green's function method (AGF) \cite{yang2018phonon, ong2015efficient} and nonequilibrium molecular dynamics simulations (NEMD) \cite{wirnsberger2015enhanced,zhu2014phonon}, as summarized in Fig. \ref{figModel} (c) and SI.
For each data point, we have 10 independent simulations, based on which the mean and standard deviations are calculated.
We first notice the negative correlation between $D_H$ and $\kappa$, $C(D_H,\kappa)=-0.95$.
For comparison between different interfaces, we define specific conductance, $\hat{\kappa}=\kappa/l$. 
Here $l$ is the Euclidean length of interface profiles. 
Interfacial conductance has been reported to scale linearly with boundary length, \cite{diao2008molecular} suggesting $\hat{\kappa}$ reduces to a constant.
However, at the atomic scale, we find instead that $\log \hat{\kappa}\propto D_H$, as shown in Fig. \ref{fig2}(a).
Note that here we only used the results for the correlation-determined interfaces in Fig. \ref{figModel} (c, left half).
The best fits in Fig. \ref{fig2}(a) for all $\Delta_m$ values are surprisingly almost the same:
$
\log\hat{\kappa}=-\gamma_D (D_H-1) + \alpha\; ,
$
$
\gamma_D=\frac{5}{2} \pm 0.02 \; ,
$
which suggests that
$$
\hat{\kappa} \propto \exp \left[ -\gamma_D (D_H-1) \right] \; .
$$
Therefore, it is the \emph{morphological irregularity}, rather than otherwise less well-defined randomness, that increases Kapitza resistance, and such effect is \emph{exponential}. 
This scaling behavior applies to wide range of $\Delta_m$.

Since only the intercept $\alpha$ differs between different mass differences, we hypothesize that $\alpha$ is a function of $\Delta_m$.
Interestingly, with the above four $\Delta_m$ cases, we find $\alpha$ depends linearly on $\log \Delta_m$, as shown in Fig. \ref{fig2}(b).
Linear regression leads to 
$$
\alpha=- \gamma_m \log \Delta_m - \gamma_0 \; ,
$$
and $\gamma_m=\frac{5}{4} \pm 0.03, \gamma_0=\frac{7}{2} \pm 0.1$,  which completes the universal scaling law,
\begin{equation} \label{eqKappa}
    \log \hat{\kappa} \approx \underbrace{-\gamma_m\log \Delta_m}_{\mathrm{mass \, diff.}} \underbrace{-\gamma_D (D_H - 1) }_{\mathrm{ structural \, irregularity}} -\gamma_0 \; .
\end{equation}
The first term on the right side describes the power-law impact of mass difference, 
the second term is the exponential effect of structural irregularity, 
and the third is a numeric prefactor. 
Increases in mass difference $\Delta_m$ and fractal dimension $D_H$ impact negatively on $\hat{\kappa}$.
Nevertheless, different from the exponential influence of $D_H$, $\Delta_m$ results in power-law impedance. 
Therefore, in terms of mass scale, the effect of mass ratio is scale-invariant.
In other words, changing mass scale will result in self-similar trend of $\kappa$.
Note that $\log \hat{\kappa} \rightarrow \infty$ as $\Delta_m \rightarrow 0$.

Before we move to the conductance quantum and structural multifractality, we test the transferrability/universality of the scaling law (Eq.\ref{eqKappa}). 
For this purpose, we use the M-W approach to generate 100 testing cases,
with varying $D_H$ and $\Delta_m$,
and 10 examples are shown in Fig. \ref{figModel} (right half, $\Delta_m=3$).
In Fig. S5, $\kappa$ of the 100 interfaces has been calculated using NEMD and is compared to the prediction from Eq. \ref{eqKappa}.
The numerical and theoretical predictions agree over a range of 6 orders of magnitude.
This verification also suggests a practical usage of Eq. \ref{eqKappa}: if $\kappa$ of one interface is known (e.g., a Poissonian interface), $\kappa$ of any random interface can be quickly estimated using Eq. \ref{eqKappa}, which also allows change in $\Delta_m$.
Further, Eq. \ref{eqKappa} proposes a metric of similarity between random interfaces.
Based on the combination of $\Delta_m$ and $D_H$, the difference between two random interfaces can be quantified.

On the other hand, however, we also note two  limitations: i) The factors of $\gamma_m=5/4$ and $\gamma_D=5/2$ have been determined numerically, and physical origins might be also interesting but yet to be determined. 
Nevertheless, these limitations occur generally to many universal laws, such as charge transport in amorphous carbons \cite{li2019charge,mandelbrot1983fractal}. 
ii) We have only considered diagonal disorder, where force potentials are assumed to be identical on the two sides of an interface. The effects of second and higher-order force constants on $\kappa$ might be further scrutinized in future. 
Nevertheless, the effects of off-diagonal disorder have been known to behave similarly to diagonal disorder in terms of wave scattering. \cite{sheng2006introduction}

Also note that our random models are at the atomic scale, which provides suitable cases to study the size effects for interface conduction. 
For instance, it has been recently measured that conductance could show quantization when characteristic length in the transverse direction falls below the coherent length of dominant phonons. \cite{rego1998quantized,gotsmann2013quantized,schwab2000measurement}
In our model systems, the feature sizes are far smaller than phonon mean free paths.
In SI, we extend the 1D coherent transport theory and show that all phononic gray media in all dimensions should have quantized conductance.
To show the possible existence of quantum conductance for interfacial transport,
in Fig. \ref{fig2} (c), we plot $\kappa$ over the universal conductance quanta, \cite{rego1998quantized}
$$
G_0=\frac{\pi^2 k_B^2}{3h} T,
$$
versus the number of interface bonds $N_B$.
Here $k_B$ and $h$ are the Boltzmann and Planck constants, $T=300$ K is the temperature.
while the non-fractals (No. 0-3) increase with $N_B$, fractal interfaces (No. 4-11) decrease with $N_B$,
and a maximum happens near $N_B=135$.
Nevertheless, all fractal interfaces exhibit number of quanta $N_0 \sim 50$.
To explain this, the local structure along an interface, which defines multifractality and mesoscale of random interfaces, is investigated in the following.


\section{Multifractality: Cantor-dust populated on geometrical fractal}
\begin{figure}[htp]
\centering
\includegraphics[width=0.99\linewidth]{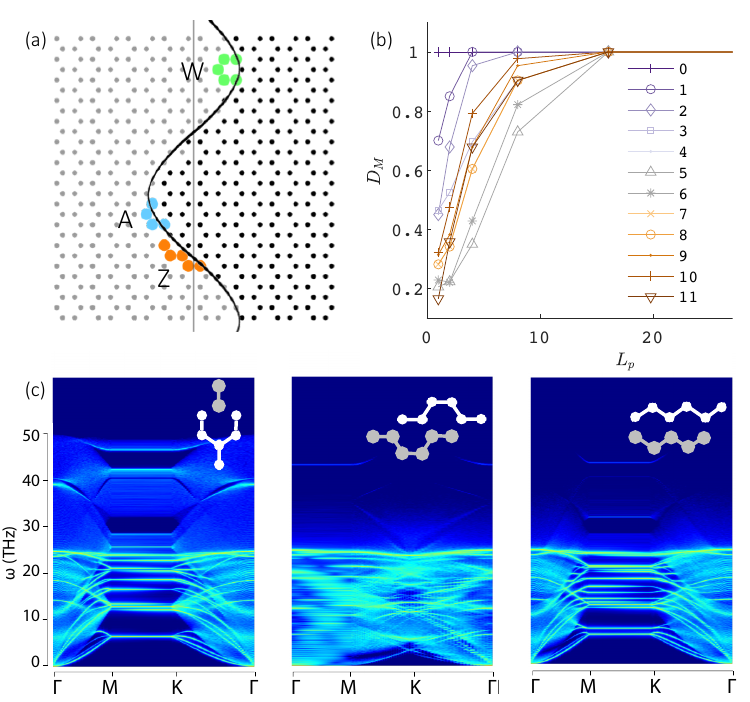}
\caption{\label{figZAW} (a) An interfacial continuum is composed at the atomic scale of zigzag (Z), armchair (A), and wrench-shaped (W) pitches. The L-system of the model interface shown in (a).
(b) Spectral energy density of the motion for interface atoms, both harmonic and anharmonic characteristics are observed to be difference between A and W/Z cases. 
(c) Minkowski dimension as a function of measurement scale. 
}
\end{figure}

As illustrated in Fig. \ref{figZAW} (a), an interface on the hexagonal lattice is composed of W-shaped, zigzag, and armchair components. 
In the limits, pristine Z-, A-, and W- interfaces have different conductance (1.6 GW/m$^2$K for Z and W interfaces, and 0.88 GW/m$^2$K for A interfaces).
This can also be seen from the spectral energy density of the interfacial atoms presented in Fig. \ref{figZAW} (b) (details in SI).
The linewidth of A-interface can be observed wider than the other two cases, serving to mixing phonon modes near the interfaces.
In other words, vibrational transport across Z$|$W-interfaces is more harmonically controlled, and A-interfaces are more anharmonically controlled.
Therefore, to fully understand the morphological effects on conductance, the distribution of structural bases will be essential information, which was however missing from the previous scaling law Eq. \ref{eqKappa}.


Towards this end, we observed such distribution \textit{per se} could also be of fractal nature.
To show this, we use the L-system (``L" for Language or Lindenmayer) to denote our interface profiles. \cite{mandelbrot1983fractal}
In the form of a sequence of characters, L-system is known as context-free grammar description of structures or rule for geometry generation.
In this work, L-system is akin to a ``DNA'' notation of our interfaces, encoded by W-, A- and Z- bases and equivalent to $\delta(x_i)$ used above. 
This L-system enables us to study the distribution of W-A-Z components using box-counting approach.
For simplicity, we only studied the distribution of single bases.
Figure \ref{figZAW} (c) shows the Minkowski dimension ($D_M$) of the Z-population for each of the model interfaces.
Pristine interface (No. 0) gives $D_M=1$ at all scales.
In contrast, random interfaces all leads to a varying dimension as measuring scale increase, although all converge to the topological dimension $D_T=1$ for larger pitches (e.g. $L_p>16$ bases for all our model interfaces).

Therefore, with regards to interfacial conduction, a random interface is a fractal populated on a geometrical fractal, which leads to the multi-fractal nature.
We realize multifractality has different definitions and distinct consequences in the literature,  such as the existence of multiple scaling laws at different scales. \cite{ivanov1999multifractality, mandelbrot1983fractal} 
In this work, we use multifractality as fractal embedded within another fractal, consistent to most mathematics literature. \cite{falconer2004fractal}
The immediate consequence of multifractality is the existence of a structural mesoscale for interficial conduction.
In our case, a correlation length $L_0\sim16$ lattice constants demarcates a regime transition, above which the distribution of basic components matters little but below which the distribution will be crucial information for interficial conductance $\kappa$.
This also explains the number of conductance quanta in the previous section, because $N_0=3 L_0 \sim 48$, where 3 is the number of modes for each degree of freedom.

\section{conclusion}
In summary, we have shown that random interfaces at the atomic scale are multifractals when vibrational transport is considered. 
The structural irregularity formulated \emph{via} the fractal dimension $D_H$ has an exponential effect on the lattice conductance.
In contrast, mass perturbation $\Delta_m$ obeys power-law scaling, and is scale-invariant and self-similar.
These two emergent laws are universally applicable to the lattice conductance of all random interfaces studied.
The correlation and distribution of interface components (e.g, W-A-Z for the planar hexagonal lattice)  also impact the conductance in a nontrivial way.
We use L-systems and their Minkowski dimension to define the structural mesoscale for interface transport, which indicates that correlation length greater than about 16 lattice constants is required for the correlated distribution of Z- and A- components to be Euclidean.
This length scale also defines the number of conductance quanta for random interfaces.
This view of random interfaces being Cantor-composites populated on geometrical fractals could be extended to charge and optical transport across random interfaces.
Nevertheless, we have not studied the interactions between the population dimension and the geometrical dimension, 
but it could be interesting for future work.

\begin{acknowledgments}
T.Z. is grateful to Dr. Xian Zhang, Dr. Arthur France-Lanord, and Dr. Kiarash Gordiz for fruitful discussions, and Dr. Ge Zhang's help with the generation of hyperuniform profiles. This work is supported by various computational resources:
(i) Comet at the Extreme Science and Engineering
Discovery Environment (XSEDE), which is supported
by National Science Foundation grant number
ACI-1548562, through allocation TG-DMR090027, and
(ii) the National Energy Research Scientific Computing
Center (NERSC), which is supported by the Office of
Science of the U.S. Department of Energy under Contract
No. DE-AC02-05CH11231.
\end{acknowledgments}

\bibliography{fractal_interface}

\end{document}